\documentclass[12pt]{article}
\usepackage{amsmath}
\usepackage{amssymb}

\newcommand{\bb}{\begin{thebibliography}{99}}
\newcommand{\eb}{\end{thebibliography}}
\newcommand{\cL}{{\cal{L}}}

\title{ EFFECTIVE ACTION AND VACUUM EXPECTATIONS IN NONLINEAR $\sigma$ MODEL }
\author{ B.A.FAYZULLAEV\\
 Department of Nuclear and Theoretical Physics,\\
 National University of Uzbekistan,\\
 Tashkent 100095, Uzbekistan}

\begin{document}

\begin{titlepage}

 \maketitle

\begin{abstract}
The equations for effective action for nonlinear $\sigma$ model are derived using DeWitt method in two forms -  for generator of vertex parts $\Gamma$ and
for generator of weakly connected parts $W$. Loop-expansion solutions to these equations are found. It is shown that vacuum expectation values
for various quantities including divergence of a N\"{o}ther current, trace of the energy-momentum tensor and so on, can be calculated by this method.
Also it is shown that vacuum expectation to the sigma-field is determined by an explicit combination of tree Green function and  classical solution.
It is shown that the limit when coupling constant tends to zero is singular one.

\end{abstract}
\vspace{.5cm}
\thispagestyle{empty}
\end{titlepage}

\section{Introduction}
One of the interesting models of quantized fields is the nonlinear
sigma-model. The range of applications of the model is very wide -
from ferromagnetism to QCD, string and superstring theories.
 Its origin is Heisenberg and Ising models of
ferromagnetism. It is clear that the model was studying from many
points of view. In \cite{nsvz} the $\sigma$ model was investigated
"as a toy model which serves sometimes as perfect theoretical
laboratory to test methods and approaches developed for solving
actual problems of actual physics". Especially interesting is that
according to Polyakov \cite{pol} there is a deeply rooted  analogy
between four dimensional Yang-Mills theories and two-dimensional
$\sigma$ models. As rule the properties of the nonlinear $\sigma-$ model were investigated on the base of the $1/N$ expansion (see, for example, \cite{nsvz}, \cite{ksen}).
We would like to present another  approach to  the studying of the nonlinear $\sigma-$ model, which is however  applicable not only  to this model but to any other quantum field models.

All the quantum properties of any model are contained in  its
effective action - quantum action. The technique developed in
\cite{fm} allows to study the effective action for any model more
deeply than other methods. In this method the effective action
$\Gamma$ is connected to the classical action $S$ by using the
DeWitt's formula (see, \cite{dw}, chapter 22). This formula leads to
some functional-derivative equations for the $\Gamma$. Such
equations for some models were solved in loop-expansion form in
\cite{fm,fbg}.

The effective action we will study is introduced as follows. Let's
to introduce the generator of all Green functions:
$$
Z[J]=\exp\{iW[J]\}=\int {\cal{D}}\varphi \exp\{i(S+J_i\varphi_i)\}.
$$
Here  the $\varphi_i$ is any field, $J_i$ is its external source,
$W[J]$ - is the generator of all connected Green's functions.
Hereafter  we will use the so-called condensed notations, for
example, $J_i\varphi_i$ means $J_i\varphi_i=\int d^4x
J(x)\varphi(x).$   Introducing so-called classical fields
\begin{equation}\label{clasfunc}
    \varphi_i=\frac{\delta W[J]}{\delta J_i}
\end{equation}
and performing following functional Legendre transformation
\begin{equation}\label{gamwjfi}
    \Gamma[\varphi]=W[J]-J_i\varphi_i
\end{equation}
we obtain the effective action $\Gamma[\varphi].$ According to
DeWitt \cite{dw}, Ch.22, the classical $S$ and quantum $\Gamma$
actions are connected as follows
\begin{equation}\label{dwrel}
    \frac{\delta\Gamma}{\delta\varphi_i}=\Lambda\frac{\delta S}{\delta\varphi_i},
\end{equation}
where the operator $\Lambda$ is constructed from connected Green
functions and functional derivatives over $\varphi_i$:
\begin{equation}\label{lambdaoper}
    \Lambda=:\exp\left\{\frac{i}{\hbar}\sum\limits_{n=2}^{\infty}\frac{(-i\hbar)^n}{n!}G_{i_1i_2\cdots
i_n}\frac{\delta^n}{\delta\varphi_{i_1}\delta\varphi_{i_2}\cdots\delta\varphi_{i_n}}\right\}:,
\end{equation}
where
$$
G_{ij}=\frac{\delta\varphi_i}{\delta J_j}=\frac{\delta^2W}{\delta
J_j \delta J_i},\qquad G_{i_1i_2\cdots i_n}=\frac{\delta^n W}{\delta
J_{i_1}\delta J_{i_2}\cdots \delta J_{i_n}}
$$
are connected (two- and $n$-point) Green functions. Commas in $\Lambda$ mean that derivatives act on r.h.s. expression only, not on $G'$s.
The
Eq.(\ref{dwrel}) connects $\Gamma$ and $n$-point Green functions
$G_{i_1i_2\cdots i_n}$, both of these are unknown quantity, so we
need in an additional relation for them. For this purpose we will
use following relation connecting the effective action $\Gamma$ and
the sources $J_i$ - so called quantum equations of motion (see
\cite{dw}):
\begin{equation}\label{delgamjot}
    \frac{\delta\Gamma}{\delta\varphi_i}=-J_i.
\end{equation}
Differentiating it over $J_j$ we will get equations we need:
$$
G^{ij}\frac{\delta^2 \Gamma}{\delta \varphi^j\delta\varphi^k}=-\delta^i_k.
$$

But we can approach to the  Eq.(\ref{delgamjot}) from another point of view -  if rewrite Eq.(\ref{dwrel}) as follows
\begin{equation}\label{eqforw}
    -J_i=\Lambda  \frac{\delta S}{\delta\varphi_i},
\end{equation}
with the $\Lambda$ operator as in Eq.(\ref{lambdaoper}) then equations for $W[J]$ will be obtained.
These equations for nonlinear $\sigma$-model are studied  in Sec.(\ref{eqw}).
In the Sec.(\ref{weakcoupling}) a very interesting conclusion about free particles in quantum vacuum following from these equations is discussed.

In this article different vacuum expectations will be calculated. Recall that a vacuum expectation of any quantum functional $ A(\hat{\varphi})$, where $\hat{\varphi}$ is a quantum field (operator valued function),
 should be read as follows \cite{dw}:
$$
\langle A(\hat{\varphi})\rangle=\frac{\langle 0, -\infty \, |\,TA(\hat{\varphi})\,|\,  0, \infty \rangle}{\langle 0, -\infty |\, 0, \infty \rangle},
$$
and it is expressed through classical functional $A(\varphi)$, where $\varphi$ is a classical field in the sense of Eq.(\ref{clasfunc}),
as follows
$$
\langle A(\hat{\varphi})\rangle=\Lambda A(\varphi).
$$
All vacuum expectations in the last sections are calculated in the absence of external sources.

\section{The model and equations}
Let's consider  following model
$$
\mathcal{L}=\frac12 (\partial_\mu\boldsymbol{\sigma})^2
$$
where
 a $N$-component scalar field $\{\sigma_a(x),\,\,a=1, 2, ..., N\}$
 is subject to the constraint
\begin{equation}\label{constr}
    \boldsymbol{\sigma}^2(x)=\sum\limits_{a=1}^N\sigma_a(x)\sigma_a(x)=\frac{N}{\gamma}.
\end{equation}
Although at first sight in the model there is no interaction but
solving the constraint Eq.(\ref{constr}) with respect to one of the
components we arrive at non-trivial interaction between remaining
components. The coefficient $\gamma$ turns out to be a coupling
constant. We can take into account this nontrivial structure of the
model by introduction of an auxiliary field - a Lagrange multiplier
- $\alpha(x)$ by the following way:
\begin{equation}\label{mod}
    \cL=\frac12 (\partial_\mu\boldsymbol{\sigma})^2-\frac12\alpha\left(\boldsymbol{\sigma}^2-\frac{N}{\gamma}\right),\qquad \boldsymbol{\sigma}=\{\sigma^a,\,a=1, 2, 3, \, ... ,\, N\},
\end{equation}
where  $\alpha(x)$ - is an auxiliary scalar field. In the condensed
notations we have for the action:
$$
S=-\frac12\sigma^a_i(\partial^2_i+\alpha_i)\sigma^a_i+\frac{N}{2\gamma}\alpha_i=-\frac12\sigma^a_iD_{ij}
\sigma^a_j +\frac{N}{2\gamma}\alpha_i,
$$
where $D_{ij}=\left(\partial^2_i+\alpha_i\right)\delta_{ij}$. Here over all the indices  summations
are implied. For obtaining of the equations for effective action we
use the following DeWitt operator \cite{fm}:
\begin{equation}
\begin{array}{l}
\displaystyle{\Lambda=:\exp\left(\frac{i}{\hbar}\sum\limits_{n=2}^{\infty}(-i\hbar)^n\sum\limits_{k=0}^n\frac{1}{(n-k)!k!}G^{a_1a_2\ldots a_k}_{i_1i_2\ldots i_kj_1j_2\ldots j_{n-k}}\frac{\delta^n}{\delta\sigma^{a_1}_{i_1}\delta\sigma^{a_2}_{i_2}\ldots\delta\sigma^{a_k}_{i_k}\delta\alpha_{j_1}\delta\alpha_{j_2}\ldots\delta\alpha_{j_{n-k}}} \right):=}\\ \\
\displaystyle{=1-i\hbar\left[\frac12\left(G^{a_1a_2}_{i_1i_2}\frac{\delta^2}{\delta\sigma^{a_1}_{i_1}\delta\sigma^{a_2}_{i_2}}+G_{j_1j_2}\frac{\delta^2}{\delta\alpha_{j_1}\delta\alpha_{j_2}}\right)+G^{a_1}_{i_1j_1}\frac{\delta^2}{\delta\alpha_{j_1}\delta\sigma^{a_1}_{i_1}}\right]+...
.}
\end{array}
\label{dwoper}
\end{equation}
Applying it to
$$
\frac{\delta\Gamma}{\delta\sigma^a_i}=\Lambda\frac{\delta
S}{\delta\sigma^a_i},\qquad
\frac{\delta\Gamma}{\delta\alpha_i}=\Lambda\frac{\delta
S}{\delta\alpha_i}
$$
 one obtains following equations  for
effective action:
\begin{equation}
\frac{\delta\Gamma}{\delta\sigma^a_i}=-D_{ij}\sigma^a_j+i\hbar
G^a_{ii},\qquad
\frac{\delta\Gamma}{\delta\alpha_i}=-\frac12\left(\boldsymbol{\sigma}^2_i-\frac{N}{\gamma}\right)+\frac{i\hbar}{2}G^{ab}_{ii}\delta_{ab}. \label{eqgamma}
\end{equation}
No summation is implied over repeated indices $i$  in these equations.
Hereafter the fields $\sigma_i^a$  and $\alpha_i$ are so-called classical fields, i.e., they are vacuum expectations of the corresponding fields in the presence of external classical sources:
\begin{equation}\label{fiwsigw}
    \sigma_i^a=\frac{\delta W}{\delta J^i_a}, \qquad \alpha_i=\frac{\delta W}{\delta \eta^i}.
\end{equation}
The following connected Green's functions arise too:
$$
G^{ab}_{ij}=\frac{\delta^2 W}{\delta J^a_i\delta J^b_j}=\frac{\delta
\sigma^a_i}{\delta J^b_j}=\frac{\delta \sigma^b_j}{\delta
J^a_i},\quad G^a_{ij}=\frac{\delta^2 W}{\delta J^a_i\delta
\eta_j}=\frac{\delta \alpha_j}{\delta J^a_i}=\frac{\delta
\sigma^a_i}{\delta \eta_j},\quad
G_{ij}=\frac{\delta^2W}{\delta\eta_i\delta\eta_j}=\frac{\delta
\alpha_i}{\delta \eta_j}=\frac{\delta \alpha_j}{\delta \eta_i}.
$$
 We should add another system of
equations which connects the $\Gamma$ and $G$'s. Applying to the
equations
\begin{equation}\label{dgamdfideta}
    \frac{\delta \Gamma}{\delta\sigma^a_i}=-J^a_i,\qquad \frac{\delta
\Gamma}{\delta\alpha_i}=-\eta_i
\end{equation}
the following operators
$$
\frac{\delta }{\delta J^b_j}=\frac{\delta\sigma^c_k}{\delta
J^b_j}\frac{\delta}{\delta\sigma^c_k}+\frac{\delta\alpha_k}{\delta
J^b_j}\frac{\delta}{\delta\alpha_k},\qquad \frac{\delta}{\delta
\eta_j}=\frac{\delta\sigma^c_k}{\delta
\eta_j}\frac{\delta}{\delta\sigma^c_k}+\frac{\delta\alpha_k}{\delta
\eta_j}\frac{\delta}{\delta\alpha_k}
$$
we obtain
\begin{equation}
\begin{array}{c}
  \displaystyle{\frac{\delta^2\Gamma}{\delta\sigma^c_k\delta\sigma^b_i}G^{ac}_{jk}+ \frac{\delta^2\Gamma}{\delta\alpha_k\delta\sigma^b_i}G^a_{jk}=-\delta^{ab}\delta_{ij},\quad  \frac{\delta^2\Gamma}{\delta\sigma^c_k\delta\alpha_i}G^{ac}_{jk}+\frac{\delta^2\Gamma}{\delta\alpha_k\delta\alpha_i}G^a_{jk}=0,}\\ \\
  \displaystyle{\frac{\delta^2\Gamma}{\delta\sigma^c_k\delta\sigma^b_j}G^{c}_{ik}+\frac{\delta^2\Gamma}{\delta\alpha_k\delta\sigma^b_j}G_{ik}=0,\qquad \frac{\delta^2\Gamma}{\delta\sigma^c_k\delta\alpha_j} G^{c}_{ik}+ \frac{\delta^2\Gamma}{\delta\alpha_k\delta\alpha_j}G_{ik}=-\delta_{ij}. }
\end{array}
\label{eqggam}
\end{equation}
Here summations over $c$ and $k$ are implied. The $a,b$ and $i,j$
indices are free ones.
 Eqs.(\ref{eqgamma}) and (\ref{eqggam})
form a closed set of equations for determination of the effective
action $\Gamma$. Recall that $\Gamma=\Gamma(\sigma,  \alpha)$  and $W=W(j, \eta).$

\section{Loop expansion for $\Gamma$}
Let's begin  with the loop expansion:
\begin{equation}
\Gamma=\sum\limits_{n=0}^{\infty}\hbar^n\Gamma_n=\Gamma_0+\hbar\Gamma_1+\hbar^2\Gamma_2+\cdots,\qquad
G=\sum\limits_{n=0}^{\infty}\hbar^nG_n=G_0+\hbar G_1+\hbar^2
G_2+\cdots. \label{eq6}
\end{equation}
Substituting this expansion into
 Eq.(\ref{eqgamma}), in the zeroth order in $\hbar$
we obtain
$$\frac{\delta\Gamma_0}{\delta\sigma^a_i}=\frac{\delta S}{\delta\sigma^a_i}, \qquad
\frac{\delta\Gamma_0}{\delta\alpha_i}=\frac{\delta
S}{\delta\alpha_i},
$$
which leads to
$$
\delta\Gamma_0=\frac{\delta
S}{\delta\sigma^a_i}\delta\sigma^a_i+\frac{\delta
S}{\delta\alpha_i}\delta\alpha_i=\delta S.
$$
So we find that
\begin{equation}
\Gamma_0=S=-\frac12\sigma^a_iD_{ij} \sigma^a_j
+\frac{N}{2\gamma}\alpha_i. \label{gamma0}
\end{equation}
The general iteration procedure is as follows: finding from
Eq.(\ref{eqgamma}) the $\Gamma_n,\quad  n=0, 1, 2, ...,$ one
substitutes it into Eq.(\ref{eqggam}), solving which one finds the
$G_n$, now from Eq.(\ref{eqgamma}) written as
\begin{equation}\label{iterG}
    \delta\Gamma_{n+1}=iG^{a}_{n\,
ii}\delta\sigma_i^a+\frac{i}{2}G^{aa}_{n\, ii}\delta\alpha_i,\qquad
n=0,\,1, \,2, \,3, \,...
\end{equation}
one can find the $\Gamma_{n+1}$, and etc. During this calculations
we will meet many times the expression like $\delta_{ii}=\delta(0)$,
which  should be  understand as regularized in some sense.
\subsection{One-loop effective action}
Substituting into   Eq.(\ref{eqggam}) the $\Gamma_0$ from (\ref{gamma0}) we obtain
equations for  $G_0$:
$$
G^{0ab}_{jk}D_{ki}+\sigma^a_iG^{0b}_{ji}=\delta_{ij}\delta^{ab},\qquad G^{0ba}_{ji}\sigma^a_i=0,
$$
$$
G^{0a}_{jk}D_{ki}+\sigma^a_iG^{0}_{ji}=0,\qquad  \sigma^a_iG^{0a}_{ji}=\delta_{ij}.
$$
In these equations  summation over $k$ is implied, no summations over $i$ and $j$, in the second and forth  equations there is summation over  $a$.
The following expressions  are solutions to these equations:
\begin{equation}\label{gam000}
G^{0ab}_{ij}=\left(D^{-1}_{ij}\delta^{ab}-\frac{\sigma^a_i\sigma^b_j}{\boldsymbol{\sigma}_i\cdot\boldsymbol{\sigma}_j} D^{-1}_{ij}\right),\qquad G^{0a}_{ij}=\frac{\sigma^a_i}{\boldsymbol{\sigma}_i\cdot\boldsymbol{\sigma}_j} \delta_{ij},\qquad G^0_{ij}=\frac{-1}{\boldsymbol{\sigma}^2_i}D_{ij}.
\end{equation}
So we find that
$$
\delta \Gamma_i=iG^{0a}_{ii}\delta\sigma^a_i+\frac{i}{2}G^{0ab}_{ii}\delta_{ab}\delta\alpha_i=i\frac{\sigma^a_i}{\boldsymbol{\sigma}_i^2} \delta_{ii}\delta\sigma^a_i+\frac{i}{2}(N-1)D^{-1}_{ii}\delta\alpha_i.
$$
It is easy to integrate this equation:
\begin{equation}\label{gamma1}
    \Gamma_1=\frac{i}{2}\mathrm{Tr} \left[\ln\boldsymbol{\sigma}^2+(N-1)\ln D\right]=\int dx \left[\ln \left(\boldsymbol{\sigma}(x)\cdot\boldsymbol{\sigma}(x)\right)+(N-1)\ln D(x,x)\right].
\end{equation}

\subsection{Two-loop effective action}

From (\ref{gamma1}) it follows that the second order partial
derivatives of the  $\Gamma_1$ are:
\begin{equation}
\frac{\delta^2\Gamma_1}{\delta\sigma^a_i\delta\sigma^c_k}=i\frac{\delta_{ik}}{\boldsymbol{\sigma}^2_i}\left(\delta^{ac}-2\frac{\sigma_i^a\sigma_i^c}{\boldsymbol{\sigma}^2_i}\right),\qquad
\frac{\delta^2\Gamma_1}{\delta\alpha_i\delta\alpha_k}=-\frac{i}{2}(N-1)D^{-2}_{ii}\delta_{ik},\qquad
\frac{\delta^2\Gamma_1}{\delta\sigma^a_i\delta\alpha_k}=0.
\label{eq10}\end{equation}
The Eqs.(\ref{eqggam}) in the first order in $\hbar$ has following form:
$$
G^{ba}_{1jk}D_{ki}+G^{b}_{1ji}\sigma^a_i=\frac{i}{\boldsymbol{\sigma}^2_i}D^{-1}_{ji}\left(\delta^{ab}-2\frac{\sigma_i^a\sigma_i^b}{\boldsymbol{\sigma}_i^2}+\frac{\sigma_i^a\sigma_j^b}{\boldsymbol{\sigma}_i\cdot \boldsymbol{\sigma}_j}\right);\quad
G^{ab}_{1ji}\sigma^b_i=-\frac{i}{2}(N-1)\frac{\sigma^a_j}{\boldsymbol{\sigma}_j\cdot \boldsymbol{\sigma}_i}D^{-2}_{ii};
$$
$$
G^{a}_{1ik}D_{ki}+G_{1ij}\sigma^a_j=-\frac{\sigma^a_j}{\boldsymbol{\sigma}_j\cdot \boldsymbol{\sigma}_i}\frac{i}{\boldsymbol{\sigma}_j^2};\qquad
G^{a}_{1ji}\sigma^a_i=\frac{i}{2}(N-1)\frac{1}{\boldsymbol{\sigma}_i\cdot \boldsymbol{\sigma}_j}D_{ji}D^{-2}_{ii}.
$$
In these equations there are no summations over $i$  and $j$ - they are free indices. In the second equation there is summation over  $b$, in the first and third equations  summation over  $k$ is implied.

According to Eq.(\ref{iterG}) our interest is only with diagonal elements of  $G^{a}_{1ij}$   and  $G^{ab}_{1ij}.$
They looks like follows  (no summation over $i$, summations over $k$ and $l$ are implied):
$$
G^{a}_{1ii}=\frac{i}{2}(N-1)D_{ik}D^{-2}_{kk}D_{kl}D^{-1}_{li}\frac{\sigma^a_l}{(\boldsymbol{\sigma}_i\cdot \boldsymbol{\sigma}_k) \,\,(\boldsymbol{\sigma}_k\cdot \boldsymbol{\sigma}_l)};
$$
$$
 G^{ab}_{1ii}\delta_{ab}=i(N-1)\left(D^{-1}_{ik}\frac{1}{\boldsymbol{\sigma}_k^2}D^{-1}_{ki}-\frac12D_{ik}D^{-2}_{kk}D_{ki}^{-1}\frac{1}{\boldsymbol{\sigma}_i\cdot \boldsymbol{\sigma}_k}\right).
$$
These lead to
$$
\delta \Gamma_2=iG^{a}_{1ii}\delta\sigma^a_i+\frac{i}{2}G^{ab}_{1ii}\delta\alpha_i=\frac{N-1}{4}\delta\left(\mathrm{Tr}\frac{1}{D\boldsymbol{\sigma}^2}\right).
$$
The effective action $\Gamma$ up to two-loop contribution is:
$$
\Gamma=-\frac12\sigma^a_iD_{ij}
\sigma^a_j +\frac{N}{2\gamma}\alpha_i+\hbar\frac{i}{2}\mathrm{Tr} \left[\ln\boldsymbol{\sigma}^2+(N-1)\ln D\right]+\hbar^2\frac{N-1}{4}\mathrm{Tr}\frac{1}{D\boldsymbol{\sigma}^2}.
$$

\section{Equations for $W$\label{eqw}}
Taking into account Eqs.(\ref{fiwsigw})  and (\ref{dgamdfideta}) we can rewrite the Eq.(\ref{eqgamma}) as  equations for $W$:
\begin{equation}\label{ww}
\begin{array}{l}
  \displaystyle{  -\partial^2\frac{\delta W}{\delta j_a(x)}-\frac{\delta W}{\delta \eta(x)}\frac{\delta W}{\delta j_a(x)}+i\hbar \frac{\delta^2W}{\delta \eta(x) \delta j_a(x) }=-j_a(x); }\\ \\
\displaystyle{\left(\frac{\delta W}{\delta j_a(x)}\frac{\delta W}{\delta j_b(x)}-i\hbar\frac{\delta^2W}{\delta j_a(x) \,\delta j_b(x) }\right)\delta_{ab}=\frac{N}{\gamma}+2\eta.}
    \end{array}
\end{equation}
The same equations we can express through classical fields $\sigma_i^a$  and $\alpha_i$ (once more using condensed notations, in the following there is no summation over $i$ but there is summation over $a$):
\begin{equation}\label{phisigmaW}\begin{array}{l}
 \displaystyle{i\hbar \frac{\delta \alpha^i}{\delta j^i_a}=i\hbar \frac{\delta \sigma^a_i}{\delta \eta_i}=\left(\partial^2+\alpha_i\right)\sigma_i^a-j^a_i=D_i\sigma_i^a-j^a_i;}\\ \\
 \displaystyle{i\hbar \frac{\delta \sigma_a^i}{\delta j^i_a}=\boldsymbol{\sigma}^2_i-\frac{N}{\gamma}-2\eta_i.}
 \end{array}\end{equation}
 Of course,  these equations may be derived using method of Eq.(\ref{eqforw}):
 $$-j^a_i=\Lambda \frac{\delta S}{\delta \sigma^a_i};\qquad -\eta_i~=~\Lambda \frac{\delta S}{\delta \alpha_i},$$
 where the $\Lambda$ is defined in Eq.(\ref{lambdaoper}). The main problem for these equations is that variational derivatives are taken in coinciding points.

\subsection{Zero-loop (classical) W}
Quasiclassical expansion for $W$ has the form
$$
W=W_0+\hbar W_1+\hbar^2 W_2+\cdots.
$$
In the same approximation the fields may be presented as follows:
$$
\sigma_a=\sigma_{0a}+\hbar \sigma_{1a}+\hbar^2\sigma_{2a}+\cdots, \qquad \alpha=\alpha_0+\hbar\alpha_1+\hbar^2\alpha_2+\cdots.
$$
From Eq.(\ref{phisigmaW}) it is easy to find the $W_0$. In this approximation we have
\begin{equation}\label{zeroloop}
    \left(\partial^2+\alpha_0\right)\boldsymbol{\sigma}_0=\mathbf{j}\quad \mbox{or},\,\, \sigma^a_{0i}=D^{-1}_{0ij}j_j^a+\tilde{\sigma}^a_i,  \qquad \boldsymbol{\sigma}^2_0=\frac{N}{\gamma}+2\eta,\qquad \alpha_0=\frac{\boldsymbol{\sigma}_0\cdot(\mathbf{j}-\partial^2\boldsymbol{\sigma}_0)}{\boldsymbol{\sigma}^2_0}.
\end{equation}
Here $\tilde{\sigma}^a_i$ means a classical solution to the equation $D_{0i}\tilde{\sigma}^a_i=0$ (instanton or meron, for example  \cite{polbel}) when $j^a_i=\eta_i=0.$
In this case the first equation in the light of the last one  may be rewritten as follows:
\begin{equation}\label{d2sd2s}
    \left(\partial^2-\frac{\boldsymbol{\tilde\sigma}_0\cdot \partial^2 \boldsymbol{\tilde\sigma}_0}{\boldsymbol{\tilde\sigma}_0^2}\right)\boldsymbol{\tilde\sigma}_0=0.
\end{equation}
Furthermore we will often consider  cases when there are no external sources.
As vacuum expectation of a scalar field the $\alpha_0$ must be constant in this case.
According to \cite{nsvz} one can put $\alpha_0=m^2$, then $m$ turns out to be an effective mass of the $\boldsymbol{\sigma}.$
In this case $\boldsymbol{\sigma}_0$ can be determined from Klein-Gordon equation
$$
\left(\partial^2+m^2\right)\boldsymbol{\sigma}_0=0,
$$
and additionally, the $\boldsymbol{\sigma}_0$ obeys the constraint $\boldsymbol{\sigma}_0^2=N/\gamma.$
The last equation and constraint are mutually agreed.

Let's calculate the $W_0:$
$$
\delta W_0=\alpha_0\delta \eta+\boldsymbol{\sigma}_0\cdot \delta \mathbf{j}=\alpha_0\boldsymbol{\sigma}_0\cdot\delta\boldsymbol{\sigma}_0+\delta(\boldsymbol{\sigma}_0\cdot \mathbf{j})-\mathbf{j}\cdot \delta \boldsymbol{\sigma}_0=
\delta(\boldsymbol{\sigma}_0\cdot \mathbf{j})+\delta\boldsymbol{\sigma}_0\left( \alpha_0\boldsymbol{\sigma}_0-\mathbf{j} \right)=
$$
$$
=\delta\left[\boldsymbol{\sigma}_0\cdot \mathbf{j}-\frac12 \boldsymbol{\sigma}_0\cdot \partial^2\boldsymbol{\sigma}_0\right]=\frac12\delta\left[\boldsymbol{\sigma}_0\cdot \mathbf{j}+2\alpha_0\eta+\frac{N}{\gamma}\alpha_0\right],
$$
from which one can conclude that
$$
W_0=\frac12\int d^4x \left(\boldsymbol{\sigma}_0\cdot \mathbf{j}+2\alpha_0\eta+\frac{N}{\gamma}\alpha_0\right).
$$
Consequently, for $\Gamma_0$ one can obtains:
$$
\Gamma_0=W_0-\int d^4x\left(\boldsymbol{\sigma}_0\cdot \mathbf{j}+\alpha_0\eta\right)=\int d^4x \left(-\frac12 \boldsymbol{\sigma}_0\cdot \partial^2\boldsymbol{\sigma}_0-\alpha_0\eta \right)=
$$
$$
=-\int d^4x \left[\frac12 \boldsymbol{\sigma}_0\cdot \partial^2\boldsymbol{\sigma}_0+\frac12\alpha_0\left(\boldsymbol{\sigma}_0^2-\frac{N}{\gamma}\right) \right].
$$
As one would expect,  the result coincides  with the classical action.

\subsection{First order quantum corrections}
Let's to calculate the first order (in $\hbar$) contributions to $\sigma_a$ and $W$.
The first order (in $\hbar$) equations:
\begin{equation}\label{phisig1}
 i \frac{\delta \alpha^i_0}{\delta j^i_a}=i \frac{\delta \sigma^a_{0i}}{\delta \eta_i}=\partial^2\sigma^a_{1i}+\alpha_{0i}\sigma_{1i}^a+\alpha_{1i}\sigma_{0i}^a=D_{0i}\sigma_{1i}^a+\alpha_{1i}\sigma_{0i}^a;\qquad
 i \frac{\delta \sigma_{0a}^i}{\delta j^i_a}=2\boldsymbol{\sigma}_{0i}\cdot \boldsymbol{\sigma}_{1i}.
 \end{equation}
 There are many relations between zero- and first order quantities.
Let's rewrite the first of  Eq.(\ref{phisig1}) as follows:
\begin{equation}\label{dsig1sig0}
     D_{0i}\sigma^a_{1i}+\alpha_{1i}\sigma^a_{0i}=i\frac{\delta \sigma^a_{0i}}{\delta \eta_i}.
\end{equation}
Using
\begin{equation}\label{delsig2}
    \sigma_{0i}^ai \frac{\delta \sigma^a_{0i}}{\delta \eta_i}=\frac12 i\frac{\delta \boldsymbol{\sigma}^2_{0i}}{\delta \eta_i}=i\delta_{ii}
\end{equation}
one can obtain:
\begin{equation}\label{alph1sigma1}
    \boldsymbol{\sigma}_{0i} \cdot D_{0i}\boldsymbol{\sigma}_{1i}+\alpha_{1i}\boldsymbol{\sigma}^2_{0i}=i\delta_{ii}.
\end{equation}
It is not so hard to solve the system of equations  (\ref{phisig1}), but difficulties emerge during this process.
Representing the Eq.(\ref{dsig1sig0}) in the following form:
\begin{equation}\label{sig1}
     \sigma^a_{1i}+D^{-1}_{0ij}\left(\alpha_{1j}\sigma^a_{0j}\right)=iD^{-1}_{0ij}\frac{\delta \sigma^a_{0j}}{\delta \eta_j}
\end{equation}
and substituting here the $\alpha_{1j}$ from Eq.(\ref{alph1sigma1}) we may find the $\sigma_1$. For the sake of simplicity let's to present the denominator of this solution in symbolic (operator) form:
\begin{equation}\label{sigma1}
\sigma_{1i}^a=-i\left[I-D_0^{-1}\left(\frac{\sigma_{0}\sigma_0}{\boldsymbol{\sigma}^2_{0}}D_0\right)\right]^{-1ab}_{il}D^{-1}_{0lj}\left\{\frac{\delta_{jj}}{\boldsymbol{\sigma}^2_{0j}}\sigma_0^b-D^{-1}_{0jk}\frac{\delta \sigma^a_{0k}}{\delta \eta_k}\right\}.
\end{equation}
Due to
$$
\left(\left[D_0^{-1}\left(\frac{\sigma_{0}\sigma_0}{\boldsymbol{\sigma}^2_{0}}D_0\right)\right]^n\right)^{ab}_{il}D^{-1}_{0lj}\frac{\delta_{jj}}{\boldsymbol{\sigma}^2_{0j}}\sigma_0^b=D^{-1}_{0ij}\frac{\delta_{jj}}{\boldsymbol{\sigma}^2_{0j}}\sigma_0^a
$$
and
$$
\left(\left[D_0^{-1}\left(\frac{\sigma_{0}\sigma_0}{\boldsymbol{\sigma}^2_{0}}D_0\right)\right]^n\right)^{ab}_{il}D^{-1}_{0lj}D^{-1}_{0jk}\frac{\delta  \alpha_{0k}}{\delta\eta_j}D^{-1}_{0ks}j^b_s=-D^{-1}_{0ij}\frac{\delta_{jj}}{\boldsymbol{\sigma}^2_{0j}}\sigma_0^a
$$
one can conclude that the r.h.s. of the Eq.(\ref{sigma1}) is of form $ 0/0.$ This ambiguity has its origin in the specific structure of Eqs.(\ref{dsig1sig0}) and(\ref{sigma01}).
Substituting the $\alpha_1$  from  Eq.(\ref{alph1sigma1}) into Eq.(\ref{dsig1sig0}) one gets:
\begin{equation}\label{orth}
    \left(\delta^{ab}-\frac{\sigma^a_{0i}\sigma^b_{0i}}{\boldsymbol{\sigma}^2_{0i}}\right)D_{0i}\sigma^b_{1i}=-i\delta_{ii}\frac{\sigma^a_{0i}}{\boldsymbol{\sigma}^2_{0i}}+i\frac{\delta \sigma^a_{0i}}{\delta \eta_i}.
\end{equation}
The operator $\delta^{ab}-\frac{\sigma^a_{0i}\sigma^b_{0i}}{\boldsymbol{\sigma}^2_{0i}}$ is projection operator onto the plane, which is orthogonal to the vector $\boldsymbol{\sigma}_0.$
So, the reason of above mentioned difficulties is that in Eq.(\ref{sigma1}) the division onto the projection operator is performed.
Here it is time to use  the second of Eqs.(\ref{gam000}). According to it we have
\begin{equation}\label{dsigde}
    \frac{\delta \sigma^a_{0i}}{\delta \eta_i}=G^a_{0ii}=\delta_{ii}\frac{\sigma^a_{0i}}{\boldsymbol{\sigma}^2_{0i}}.
\end{equation}
Consequently,  Eq.(\ref{orth}) may be rewritten   as follows (the subscripts $\perp$ and $\parallel$ hereafter mean "orthogonal to $\boldsymbol{\sigma}_0$" and "parallel to $\boldsymbol{\sigma}_0$", respectively):
$$
\left(D_{0i}\boldsymbol{\sigma}_{1i}\right)_\perp=0.
$$
So due to Eqs.(\ref{dsig1sig0}) and (\ref{dsigde}) we may write down
\begin{equation}\label{dsig1}
    D_0\boldsymbol{\sigma}_{1i}=\left(-\alpha_{1i}+i\frac{\delta_{ii}}{\boldsymbol{\sigma_{0i}}^2}\right)\boldsymbol{\sigma_{0i}}.
\end{equation}
Hence it is shown that
\begin{equation}\label{sigma01}
    \sigma_{1i}^a=D^{-1}_{0ij}\left[\left(-\alpha_{1j}+i\frac{\delta_{jj}}{\boldsymbol{\sigma_{0j}}^2}\right)\sigma^a_{0j}\right].
\end{equation}
Now we should to calculate the $\alpha_1$. From the last term in the Eq.(\ref{phisig1}) and the second term in the Eq.(\ref{zeroloop}) one obtains
\begin{equation}\label{phi0phi1djd}
     2\boldsymbol{\sigma}_{0i}\cdot \boldsymbol{\sigma}_{1i}=i \frac{\delta \sigma_{0a}^i}{\delta j^i_a}=iND^{-1}_{0ii}-iD^{-1}_{0ik} \frac{\delta \alpha_{0k} }{\delta j^a_i}  D^{-1}_{0kl}j^a_l.
\end{equation}
But according to the first of Eqs.(\ref{gam000}) we have:
$$
\frac{\delta \sigma_{0a}^i}{\delta j^i_a}=G_{0ii}^{ab}\delta_{ab}=(N-1)D^{-1}_{0ii},
$$
which means that
\begin{equation}\label{sig0sig1}
\boldsymbol{\sigma}_{0i}\cdot \boldsymbol{\sigma}_{1i}=\frac{i}{2}(N-1)D^{-1}_{0ii}.
\end{equation}
Using this result and Eq.(\ref{sigma1}) one gets
\begin{equation}\label{sig0dsig0}
    \sigma^a_{0i}D^{-1}_{0ij}\left(\alpha_{1j}\sigma^a_{0j}\right)=i\delta_{ii}\sigma^a_{0i}D^{-1}_{0ij}\left(\frac{\sigma^a_{0j}}{\boldsymbol{\sigma}_{0j}^2}\right)-\frac{i}{2}(N-1)D^{-1}_{0ii}.
\end{equation}
  We will solve this integral equation only in the source free approximation $\mathbf{j}=\eta=0$ (because we will use this case in subsequent calculations). In this case the $\alpha$  as vacuum expectation of
scalar field must be a constant, so we have:
\begin{equation}\label{alfa1}
    \alpha_{1}=i\frac{\gamma}{N}\delta_{ii}-\frac{i(N-1)}{2}\frac{1}{\boldsymbol{\tilde\sigma_0}\cdot D^{-1}_0\boldsymbol{\tilde\sigma_0}}D^{-1}_{0ii}.
\end{equation}
Substituting this expression into Eq.(\ref{sigma01}) we get:
\begin{equation}\label{sigma1i}
    \sigma_{1i}^a=\frac{i(N-1)}{2}\frac{1}{\boldsymbol{\tilde\sigma_0}\cdot D^{-1}_0\boldsymbol{\tilde\sigma_0}}D^{-1}_{0kk}\cdot D^{-1}_{0ij}\tilde\sigma_{0j}^a.
\end{equation}
It is obvious that this formula is compatible with Eq.(\ref{sig0sig1}).
In the last two formulas under $\tilde\sigma_0$ we should understand a solution of the equation (\ref{d2sd2s}).
The surprising thing is that quantum corrections to vacuum expectations are expressed in terms of classical solutions!

In later sections calculations  we need in the following combination of fields:
\begin{equation}\label{sig0d1sig1}
\boldsymbol{\tilde\sigma_{0i}}\cdot D_{0i}\boldsymbol{\tilde\sigma_{1i}}=i\frac{N-1}{2}\frac{\boldsymbol{\tilde\sigma_0}^2}{\boldsymbol{\tilde\sigma_0}\cdot D^{-1}_0\boldsymbol{\tilde\sigma_0}}D^{-1}_{0kk}=i\frac{N}{\gamma}\frac{N-1}{2}\frac{1}{\boldsymbol{\tilde\sigma_0}\cdot D^{-1}_0\boldsymbol{\tilde\sigma_0}}D^{-1}_{0kk}.
\end{equation}
Emphasize once more, that three last formulas (\ref{alfa1}), (\ref{sigma1i})  and (\ref{sig0d1sig1}) are derived for  source free  case.
In all the denominators of these formulas  the expression $\boldsymbol{\tilde\sigma_0}\cdot D^{-1}_0\boldsymbol{\tilde\sigma_0}$
 presents.
 It depends only on classical solutions for the sigma-model and represents a numeric quantity. It is  easy to verify that this functional is  proportional to $N/(m^2\gamma).$
 Let's denote it as follows
\begin{equation}\label{sigdsig}
    \boldsymbol{\tilde\sigma_0}\cdot D^{-1}_0\boldsymbol{\tilde\sigma_0}=\int d^{\,d}x\, d^{\,d}y \,\,\tilde\sigma^a(x) D^{-1}_0(x, y)\tilde\sigma^a(y)=\frac{N}{m^2\gamma}C[\tilde\sigma].
\end{equation}
Taking into account (\ref{sigma1i}) we may present the full vacuum expectation of the $\sigma-$field  as
$$
\sigma^a_{1i}=\left(\delta_{ij}+i\hbar\frac{N-1}{2}\frac{\gamma}{NC[\tilde\sigma]}D^{-1}_{0kk} D^{-1}_{ij}\right)\tilde\sigma^a_{0j},
$$
 Detailed calculation of the constant $C[\tilde\sigma]$  for instanton  solution to nonlinear sigma-model (in the case $N=3,\,\,d=2$) will be presented in the other publication.

\section{About weak coupling limit\label{weakcoupling}}
Let's consider scalar theories with interactions $\lambda \varphi^3$  and $\lambda \varphi^4.$ Classical actions for these theories are
$$
S=-\frac12 \varphi_i(\partial^2+m^2)\varphi_i-\lambda \varphi_i^3=-\frac12 \varphi_i K^{-1}_{ij}\varphi_j-\lambda \varphi_i^3
$$
and
$$
S=-\frac12 \varphi_i(\partial^2+m^2)_{ij}\varphi_j-\lambda \varphi^4=-\frac12 \varphi_i K^{-1}_{ij}\varphi_j-\lambda \varphi_i^4.
$$
Applying to these classical actions Eq.(\ref{eqforw})  with $\Lambda$  from (\ref{lambdaoper}) one can obtains equations for $W:$
$$
\frac{i\lambda\hbar}{2}\frac{\delta^2 W}{\delta J_i^2}-\frac{\lambda}{2}\left(\frac{\delta W}{\delta J_i}\right)^2-K_{ij}^{-1}\frac{\delta W}{\delta J_j}+J_i=0;
$$
and
$$
\frac{\lambda}{6}\hbar^2\frac{\delta^3W}{\delta J_i^3}+\frac{i\lambda\hbar}{2}\frac{\delta^2W}{\delta J_i^2}\frac{\delta W}{\delta J_i}-\frac{\lambda}{6}\left(\frac{\delta W}{\delta J_i}\right)^3-K^{-1}_{ij}\frac{\delta W}{\delta J_j}+J_i=0.
$$
The main difficulty concerned with these equations is coincidence of arguments in variational derivatives.
But especially surprising point expects us in weak coupling limit  $\lambda \rightarrow 0$  -
  these equations (as any other equations for effective action too) are ones with small parameter in front of higher derivative terms.
  As it is well known for such type of  (linear at least) equations the limit  $\lambda \rightarrow 0$  is singular. That is, if we put $\lambda=0$ in these equations
  then their solution $W_0=\frac12 J_i K_{ij}J_j$ can not be considered as any approximation to the exact $W$ because
  the exact $W$ is singular at $\lambda=0.$
As is well known, the series over small $\lambda$ which can be derived from these equations may be  asymptotic one only.

  This means that the notion of free particle very often used in textbooks on quantum field theory  needs to be revised.

The same situation we have in the $\sigma$-model equations (\ref{ww})  and  (\ref{phisigmaW}) - second equations in these systems in the weak coupling limit $\gamma \rightarrow 0$
turns to have  $N =O(\gamma)$ form, i.e., $\gamma=0$ is singular point for the $\sigma$-model too.

\section{Conservation law}\label{conslaw}
Because the Lagrangian (\ref{mod}) is invariant under $O(N)$ group of transformations
$$
\sigma_a'(x)=O_{ab}\sigma_b(x)\simeq \left(\delta_{ab}+i\delta \alpha A_{ab}\right)\sigma_j,\qquad A_{ab}=-A_{ba},\qquad a, b=1, 2, ..., N,
$$
there is a subsequent  N\"{o}ther current
$$
J_\mu(x)=-i\partial_\mu\sigma_a(x)  A_{ab}\sigma_b(x).
$$
Its conservation is consequence of classical equations of motion  $\partial^2\sigma_a+\alpha\sigma_a=0$  and antisymmetry of generators $A_{ab}$:
$$
\partial_\mu J^\mu=-i\partial^2\sigma_a A_{ab}\sigma_b=i\alpha \sigma_aA_{ab}\sigma_b=0.
$$
In quantum case we have
\begin{equation}\label{conserv}
\begin{array}{l}
\displaystyle{\left<\partial_\mu J^\mu(x)\right>=-i \left<\partial^2\sigma_a(x)  A_{ab}\sigma_b(x)\right>=-i\Lambda \left(\partial^2\sigma_a(x)  A_{ab}\sigma_b(x)\right)=}\\ \\
\displaystyle{=-i\partial^2\sigma_a(x) A_{ab}\sigma_b(x)-\frac{\hbar}{2}\mathrm{Tr} \left[\partial^2_x G_{ab}(x, y)A_{ab}\right]\Big|_{y=x}.}
\end{array}\end{equation}
Let's rewrite this equation up to one loop contributions:
\begin{equation}\label{concquant}
\begin{array}{c}
    \left<\partial_\mu J^\mu(x)\right> \simeq-i\partial^2\sigma^0_a(x) A_{ab}\sigma^0_b(x)+  \\  \\
+i\hbar\left[-\partial^2\sigma^0_a(x) A_{ab}\sigma^1_b(x)-\partial^2\sigma^1_a(x) A_{ab}\sigma^0_b(x)+\frac{i}{2}\mathrm{Tr} \left[\partial^2_x G^0_{ab}(x, y)A_{ab}\right]\Big|_{y=x}\right].
\end{array}\end{equation}
In what follows we will consider no external source case  (i.e., only $\mathbf{j}=\eta=0$ case formulas of the Sec.\ref{eqw}  will be used).

The classical contribution (the first term) as it had been shown, vanishes. Let's consider the one-loop contribution.
First two terms are:
$$
-\partial^2\sigma_{0a}A_{ab}\sigma_{1b}=\alpha_0\sigma_{0a}A_{ab}\sigma_{1b};
$$
$$
-\partial^2\sigma_{1a} A_{ab}\sigma_{0b}=\left(\alpha_1\sigma_{0a}+\alpha_0\sigma_{1a}\right)A_{ab}\sigma_{0b}=\alpha_0\sigma_{1a}A_{ab}\sigma_{0b}.
$$
 From antisymmetry of $A_{ab}$ it follows that their sum is equal to zero:
$$
-\partial^2\sigma^0_a(x) A_{ab}\sigma^1_b(x)-\partial^2\sigma^1_a(x) A_{ab}\sigma^0_b(x)=\alpha_0\left(\sigma_{0a}A_{ab}\sigma_{1b}+\sigma_{1a}A_{ab}\sigma_{0b}\right)=0.
$$
For treatment of the last term in Eq.(\ref{concquant}) we will use the first expression in Eq.(\ref{gam000}): it is obviously symmetric in its $(a,b)$  indices, so
due to antisymmetry of $A_{ab}$ we have
$$
\mathrm{Tr} \left[\partial^2_x G^0_{ab}(x, y)A_{ab}\right]\Big|_{y=x}=0.
$$
So it has been proved that in one-loop approximation
$$
\left<\partial_\mu J^\mu(x)\right>=0.
$$
It is well known that if there is no anomaly in one-loop level there is no anomaly in general \cite{sf}.
The $O(N)$ symmetry of the model is exact.

\section{Trace of the energy-momentum tensor\label{traceemt}}
Classical canonical energy-momentum tensor for nonlinear
$\sigma$-model has the form:
\begin{equation}\label{canontensor}
    T^\mu_\nu=\frac{\partial \mathcal{L}}{\partial \partial_\mu\sigma^a_i}\partial_\nu\sigma^a_i-\delta^\mu_\nu\mathcal{L}=\partial^\mu\boldsymbol{\sigma}\partial_\nu\boldsymbol{\sigma}-\frac12\delta^\mu_\nu\left[(\partial_\lambda\boldsymbol{\sigma})^2-\alpha\left(\boldsymbol{\sigma}^2-\frac{N}{\gamma}\right)\right].
\end{equation}
Because $\sigma^a$ is scalar field in $x-$space this canonical tensor coincides with symmetrical one of Belinfante.
Trace of this tensor is (in $d$-dimensional space-time):
\begin{equation}\label{classictrace}
    T^\mu_\mu=\left(1-\frac{d}{2}\right)(\partial_\lambda\boldsymbol{\sigma})^2+\frac{d}{2}\alpha\left(\boldsymbol{\sigma}^2-\frac{N}{\gamma}\right).
\end{equation}
In $d=2$ space-time we have (taking into account the classical
equations of motion)
$$
 T^\mu_\mu=0.
$$
According to DeWitt  the vacuum
expectation value of this trace is
\begin{equation}\label{quantumtrace}\begin{array}{c}
  \displaystyle{  \langle T^\mu_\mu\rangle=\Lambda T^\mu_\mu=\left(1-\frac{d}{2}\right)(\partial_\lambda\boldsymbol{\sigma})^2+\frac{d}{2}\alpha\left(\boldsymbol{\sigma}^2-\frac{N}{\gamma}\right)-\frac{i\hbar d}{2}\alpha G^{ab}_{ii}\delta_{ab}-i\hbar d \sigma^a_i G^a_{ii}+}\\ \\
\displaystyle{+i\hbar\left(\frac{d}{2}-1\right)\left.\partial^x_\mu\partial_y^\mu G^{ab}(x, y)\right|_{x=y}\delta^{ab}+
\frac{\hbar^2d}{4}G^{ijk}_{ab}\delta^{ab}\delta_{ik}\delta_{jk},}
\end{array}\end{equation}
where summations over all the indices are implied. Here  $G^{ab}_{ijk}$  is three-point connected Green function:
$$
G^{ijk}_{ab}=\frac{\delta^3 W}{\delta j^a_i\delta j^b_j\delta\eta_k}\Bigg|_{\mathbf{j}=0,\,\eta=0}.
$$
Of course, the fields $\sigma^a_i$ and $\alpha_i$  in Eqs.(\ref{classictrace})  and (\ref{quantumtrace}) has different meaning - in (\ref{classictrace}) they are classical functions,
but in  (\ref{quantumtrace}) they are vacuum expectations of corresponding quantum fields. This means the first four terms in (\ref{quantumtrace}) should be considered (in contrast to (\ref{classictrace})) as dependent on $\hbar.$

In the zero-loop approximation (putting in (\ref{zeroloop}) $\mathbf{j}=0,\,\,\eta=0$) we obtain
\begin{equation}\label{tmn}
    \left.\langle T^\mu_\mu\rangle\right|_{\hbar=0}=\frac{2-d}{2}\partial^\mu\boldsymbol{\sigma}_0\partial_\mu\boldsymbol{\sigma}_0=\frac{2-d}{2}\left[\partial_\mu\left(\boldsymbol{\sigma}_0\partial^\mu\boldsymbol{\sigma}_0\right)+\alpha_0\boldsymbol{\sigma}_0^2\right]=\frac{2-d}{2}\frac{N}{\gamma}m^2 ,
\end{equation}
due to $\partial_\mu\left(\boldsymbol{\sigma}_0\partial^\mu\boldsymbol{\sigma}_0\right)=\frac12 \partial_\mu\left(\partial^\mu \boldsymbol{\sigma}^2_0\right)=\partial^2\eta=0.$
The $\alpha_0$  as vacuum expectation of a (source free) scalar field may be constant only. In \cite{nsvz} it was suggested to take $\alpha_0=m^2$, where $m$ turns out to be an effective mass of  the $\boldsymbol{\sigma}$ field.
Of course, in $d=2$ space-time we have result consistent with classical one (on-shell)
$$
\left.\langle T^\mu_\mu\rangle\right|_{\hbar=0}=0.
$$
In the one-loop approximation we have
$$
\langle T^\mu_\mu\rangle_{1}=(2-d)\partial^\mu\boldsymbol{\sigma}_0\cdot \partial_\mu\boldsymbol{\sigma}_1+\alpha_0 d \boldsymbol{\sigma}_0\cdot\boldsymbol{\sigma}_1-\frac{i d}{2}\alpha_0G_{0ii}^{ab}\delta_{ab}-id\sigma_{0i}^aG_{0ii}^a+\frac{i}{2}(d-2)\left.\partial^x_\mu\partial_y^\mu G^{ab}_0(x, y)\right|_{x=y}\delta^{ab}.
$$
Acting on Eq.(\ref{sig0sig1})  by operator $\partial^2$ and equating the result to zero one can obtains:
\begin{equation}\label{dvasig0}
    2\partial_\mu\boldsymbol{\sigma}_{0i}\cdot \partial^\mu\boldsymbol{\sigma}_{1i}=2m^2\boldsymbol{\sigma}_{0i}\cdot \boldsymbol{\sigma}_{1i}-\boldsymbol{\sigma}_{0i}\cdot D_0\boldsymbol{\sigma}_{1i}=i\frac{N-1}{2}D^{-1}_{oii}\left[2m^2-\frac{N}{\gamma}\frac{1}{\boldsymbol{\sigma_0}\cdot D^{-1}_0\boldsymbol{\sigma_0}}\right]
\end{equation}
So the full one-loop contribution to $\langle T^\mu_\mu\rangle$  is (Eq.(\ref{sigdsig})  was taken into account too):
$$
\langle T^\mu_\mu\rangle_1=-i\delta_{ii}d+i\frac{2-d}{4}(N-1)\left[\left(2m^2-\frac{m^2}{C[\tilde\sigma]}\right)D^{-1}_{0ii} - 2\left.\partial^x_\mu\partial_y^\mu D^{-1}_0(x, y)\right|_{x=y}\right],
$$
where
$$
 D_0^{-1}(x, y)=\int \frac{d^d p}{(2\pi)^d}\frac{e^{-ip(x-y)}}{m^2-p^2}.
$$
Calculation of this expression in two dimensional space-time using the dimensional regularization method ($d=2-2\varepsilon$) gives:
$$
\langle T^\mu_\mu\rangle_{1}=\frac{N-1}{8\pi C[\tilde\sigma]}m^2.
$$
Non vanishing trace of the energy-momentum tensor presents conformal anomaly and it is not surprising that it is proportional to $m^2.$ As it is well known, originally massless $\sigma$-field due to spontaneous symmetry breaking acquires a mass.
 Then the density of vacuum energy will be
$$
\varepsilon_{vac}=\frac{1}{d}\hbar\langle T^\mu_\mu\rangle_1=\hbar\frac{N-1}{16\pi C[\tilde\sigma]}m^2.
$$
In the dimensional regularization method the answer is convergent, but if one  use  the momentum cutting method then due to the term $\delta_{ii}=\delta(0)$
a quadratic divergence proportional to $M^2$ (cutoff parameter) would be  appeared.
It appears that  the vacuum energy is determined by classical solutions only. In the case $N=3,\,\,d=2$  nothing but instanton contributes to the vacuum energy.

\section{Factorization theorems\label{vaccon}}
It is interesting to calculate following vacuum condensates $\langle\boldsymbol{\sigma}^{2k}\rangle,$ where $k=1,, 2, 3, ...$
Let's begin with the $k=1:$
\begin{equation}\label{sig2n}
    \langle\boldsymbol{\sigma}^2_i\rangle=\boldsymbol{\sigma}_{i}^2-i\hbar G^{ab}(x,x)\delta_{ab}=\boldsymbol{\sigma}_{0i}^2+2\hbar\boldsymbol{\sigma}_{0i}\cdot \boldsymbol{\sigma}_{1i}-i\hbar(N-1) D^{-1}_{0ii}+\cdots.
\end{equation}
In the tree approximation we have
$$
\langle\boldsymbol{\sigma}^2\rangle_0=\boldsymbol{\sigma}^2_0=\frac{N}{\gamma}.
$$
 The second and third terms in Eq.(\ref{sig2n}) cancel out due to (\ref{sig0sig1}). So the one-loop correction to the $\boldsymbol{\sigma}^2$ vanishes and we have
\begin{equation}\label{sigkvad}
    \langle\boldsymbol{\sigma}^2_i\rangle=\frac{N}{\gamma}+O(\hbar^2).
\end{equation}
In the same way the vacuum expectation of $\boldsymbol{\sigma}^{2k}_i$  may be calculated:
$$
\langle\boldsymbol{\sigma}^{2k}_i\rangle=\boldsymbol{\sigma}^{2k}_i-\frac{i\hbar}{2}G^{jl}_{ab}\frac{\delta^2}{\delta\sigma^j_a\delta\sigma^l_b}\boldsymbol{\sigma}^{2k}_i=
\boldsymbol{\sigma}^{2k}_{0i}+2k\hbar\boldsymbol{\sigma}_{0i}^{2k-2}\boldsymbol{\sigma}_{0i}\cdot\boldsymbol{\sigma}_{1i}-ik\hbar G^{ab}_{0ii}\delta_{ab}\boldsymbol{\sigma}^{2k-2}_{0i}+\cdots=
$$
$$
=\boldsymbol{\sigma}^{2k}_{0i}+O(\hbar^2)=
\left(\frac{N}{\gamma}\right)^k+O(\hbar^2).
$$
Within  the one-loop accuracy (at least) one may write:
$$
\langle\boldsymbol{\sigma}^{2k}_i\rangle=\left(\frac{N}{\gamma}+O(\hbar^2)\right)^k=\langle\boldsymbol{\sigma}^2_i\rangle^k, \qquad k\geq 1.
$$
Let's calculate  the following vacuum condensate:
$$
\langle\partial^\mu\boldsymbol{\sigma}\cdot \partial_\mu\boldsymbol{\sigma}\rangle=\partial^\mu\boldsymbol{\sigma}\cdot \partial_\mu\boldsymbol{\sigma}-i\hbar\left.\partial^x_\mu\partial_y^\mu G^{ab}(x, y)\right|_{x=y}\delta^{ab}.
$$
In the zero-loop approximation one obtains (cf. derivation of the Eq.(\ref{tmn}))
$$
\left.\langle\partial^\mu\boldsymbol{\sigma}\cdot \partial_\mu\boldsymbol{\sigma}\rangle\right|_{\hbar=0}=\partial^\mu\boldsymbol{\sigma}_0\cdot \partial_\mu\boldsymbol{\sigma}_0=\frac{N}{\gamma}m^2.
$$
One-loop contribution to this condensate is:
$$
\langle\partial^\mu\boldsymbol{\sigma}_i\cdot \partial_\mu\boldsymbol{\sigma}_i\rangle_1=2\partial^\mu\boldsymbol{\sigma}_{0i}\cdot \partial_\mu\boldsymbol{\sigma}_{1i}-i\left.\partial^x_\mu\partial_y^\mu G^{ab}_0(x, y)\right|_{x=y}\delta^{ab}.
$$
Taking into account Eq.(\ref{dvasig0}) we have
\begin{equation}\label{onedsds}\begin{array}{c}
  \displaystyle{  \langle\partial^\mu\boldsymbol{\sigma}_i\cdot \partial_\mu\boldsymbol{\sigma}_i\rangle_1=i\frac{N-1}{2}D^{-1}_{oii}\left[2m^2-\frac{m^2}{C[\tilde\sigma]}\right]-i(N-1)\left.\partial^x_\mu\partial_y^\mu D^{-1}_0(x, y)\right|_{x=y}}\Rightarrow\\ \\
\displaystyle{=  \frac{i}{2}(N-1)\int \frac{d^dp}{(2\pi)^d} \frac{2m^2-2p^2-m^2/ C[\tilde\sigma]}{m^2-p^2}=\frac{N-1}{8\pi C[\tilde\sigma]}m^2N_\varepsilon,\quad N_\varepsilon=\frac{1}{\varepsilon}-\gamma_E-\ln(4\pi).}
\end{array}\end{equation}
So it is obtained that
\begin{equation}\label{renconst}
\langle\partial^\mu\boldsymbol{\sigma}_i\cdot \partial_\mu\boldsymbol{\sigma}_i\rangle=\frac{N}{\gamma}m^2+\hbar\frac{N-1}{8\pi C[\tilde\sigma]}m^2N_\varepsilon+O(\hbar^2)=\partial^\mu\boldsymbol{\sigma}_0\cdot \partial_\mu\boldsymbol{\sigma}_0\left(1+\hbar\frac{N-1}{8\pi}\frac{\gamma}{N  C[\tilde\sigma]}N_\varepsilon+\cdots\right).
\end{equation}

It is interesting to calculate  $\langle(\partial^\mu\boldsymbol{\sigma}\cdot \partial_\mu\boldsymbol{\sigma})^k\rangle$ \cite{nsvz}. Acting the same way as above
we obtain:
$$
\langle(\partial^\mu\boldsymbol{\sigma}_i\cdot \partial_\mu\boldsymbol{\sigma}_i)^k\rangle=(\partial^\mu\boldsymbol{\sigma}_i\cdot \partial_\mu\boldsymbol{\sigma}_i)^k-\frac{i\hbar}{2}G^{lk}_{ab}\frac{\delta^2}{\delta\sigma^l_a\delta\sigma^k_b}(\partial^\mu\boldsymbol{\sigma}_i\cdot \partial_\mu\boldsymbol{\sigma}_i)^k+\cdots=
$$
$$
=(\partial^\mu\boldsymbol{\sigma}_{0i}\cdot \partial_\mu\boldsymbol{\sigma}_{0i})^k+2k\hbar(\partial^\mu\boldsymbol{\sigma}_{0i}\cdot \partial_\mu\boldsymbol{\sigma}_{0i})^{k-1}\partial^\nu\boldsymbol{\sigma}_{0i}\cdot \partial_\nu\boldsymbol{\sigma}_{1i}-ik\hbar \partial^\mu_l\partial_\mu^kG^{ab}_{lk}\delta_{lk}\delta^{ab}(\partial^\nu\boldsymbol{\sigma}_{0i}\cdot \partial_\nu\boldsymbol{\sigma}_{0i})^{k-1}-
$$
$$
-2i\hbar k(k-1)G^{ab}_{lk}\cdot\partial_\mu^i\delta_{li}\partial_\nu^i\delta_{ki} \cdot\partial^\mu_i\sigma_{0i}^a\cdot\partial^\nu_i\sigma_{0i}^b(\partial^\lambda\boldsymbol{\sigma}_{0i}\cdot \partial_\lambda\boldsymbol{\sigma}_{0i})^{k-2}+O(\hbar^2).
$$
The last term requires an accurate processing. The following chain of formulas solves the problem (no summation over index $i$, summations over $k$  and $l$ carrying out):
$$
G^{ab}_{lk}\cdot\partial_\mu^i\delta_{li}\partial_\nu^i\delta_{ki} \cdot\partial^\mu_i\sigma_{0i}^a\cdot\partial^\nu_i\sigma_{0i}^b=\left(\partial_\mu\sigma^a_{0i}\partial_\nu\sigma^a_{0i} D^{-1}_{lk}-\frac{(\sigma_{0l}^a\partial_\mu\sigma_{0i}^a)(\sigma_{0k}^b\partial_\nu\sigma_{0i}^b)}{\boldsymbol{\sigma}_{0l}\cdot\boldsymbol{\sigma}_{0k}}D^{-1}_{lk}\right)\partial_\mu^i\delta_{li}\partial_\nu^i\delta_{ki}=
$$
$$
=\delta_{li}\delta_{ki}\partial_\nu^k\partial_\mu^l\left(\partial_\mu\sigma^a_{0i}\partial_\nu\sigma^a_{0i} D^{-1}_{lk}-\frac{(\sigma_{0l}^a\partial_\mu\sigma_{0i}^a)(\sigma_{0k}^b\partial_\nu\sigma_{0i}^b)}{\boldsymbol{\sigma}_{0l}\cdot\boldsymbol{\sigma}_{0k}}D^{-1}_{lk}\right)=\left.\partial_\mu\sigma^a_{0i}\partial_\nu\sigma^a_{0i} \partial^\mu_l\partial^\nu_kD^{-1}_{lk}\right|_{l=k=i}-
$$
$$
-\left.\frac{\left(\partial^\mu\boldsymbol\sigma_{0i}\cdot\partial_\mu\boldsymbol\sigma_{0i}\right)^2}{\boldsymbol\sigma_{0i}^2}D^{-1}_{lk}\right|_{l=k=i}=\left.\frac{1}{d}(\partial^\mu\boldsymbol\sigma_{0i}\cdot\partial_\mu\boldsymbol\sigma_{0i})\partial_\nu^l\partial^\nu_kD^{-1}_{kl}\right|_{k=l=i}-\frac{N}{\gamma}m^4D^{-1}_{ii}=
$$
$$
=\left.\frac{Nm^2}{\gamma}\left(\frac{1}{d}\partial_\nu^l\partial^\nu_kD^{-1}_{kl}\right|_{k=l=i}-m^2D^{-1}_{ii}\right).
$$
During this calculation all terms proportional to $\boldsymbol\sigma_{0i}\cdot\partial_\mu\boldsymbol\sigma_{0i}=\frac12\partial_\mu \boldsymbol\sigma_{0i}^2=0$ are omitted.
Then we have:
$$
\langle(\partial^\mu\boldsymbol{\sigma}_i\cdot \partial_\mu\boldsymbol{\sigma}_i)^k\rangle=\left(\frac{Nm^2}{\gamma}\right)^k+k\hbar\left(\frac{Nm^2}{\gamma}\right)^{k-1}\frac{m^2}{8\pi}\left[\frac{N-1}{ C[\tilde\sigma]}-2(k-1)\right]N_\varepsilon+O(\hbar^2).
$$
Comparing this result with Eq.(\ref{renconst}) we may conclude that appearance of the last term in brackets means absence of factorization at one-loop level already:
$$
\langle(\partial^\mu\boldsymbol{\sigma}_i\cdot \partial_\mu\boldsymbol{\sigma}_i)^k\rangle\neq\langle\partial^\mu\boldsymbol{\sigma}_i\cdot \partial_\mu\boldsymbol{\sigma}_i\rangle^k,\qquad k\geq 2.
$$
But in the leading $N$ approximation the factorization is held.

It is instructive to calculate the following condensate (in one-loop approximation):
$$\begin{array}{c}
\langle (\boldsymbol\sigma\cdot\partial_\mu\boldsymbol\sigma)^2\rangle=(\boldsymbol\sigma\cdot\partial_\mu\boldsymbol\sigma)^2-i\hbar G^{ab}_{lk}\left[\left(\delta_{li}\partial_\mu^i\sigma_{0i}^a+\sigma_{0i}^a\partial_\mu^i\delta_{li}\right)\left(\delta_{ki}\partial_\mu^i\sigma_{0i}^b+\sigma_{0i}^b\partial_\mu^i\delta_{ki}\right)+\right.\\ \\
\left.+\boldsymbol\sigma_{0i}\cdot\partial_\mu\boldsymbol\sigma_{0i}(\delta_{ki}\partial_\mu^i\delta_{li}+\delta_{li}\partial_\mu^i\delta_{ki})\right].
\end{array}$$
The last term vanishes due to $\boldsymbol\sigma_{0i}\cdot\partial_\mu\boldsymbol\sigma_{0i}=0,$ contribution of the first term begins at two-loop level due to the same reason:
$$
\boldsymbol\sigma\cdot\partial_\mu\boldsymbol\sigma=\boldsymbol\sigma_0\cdot\partial_\mu\boldsymbol\sigma_0+\hbar\left(\boldsymbol\sigma_1\cdot\partial_\mu\boldsymbol\sigma_0+\boldsymbol\sigma_0\cdot\partial_\mu\boldsymbol\sigma_1\right)+O(\hbar^2)=\hbar\left(\boldsymbol\sigma_1\cdot\partial_\mu\boldsymbol\sigma_0+\boldsymbol\sigma_0\cdot\partial_\mu\boldsymbol\sigma_1\right)+O(\hbar^2).
$$
Due to this circumstance already it is seen that  factorization like $\langle (\boldsymbol\sigma\cdot\partial_\mu\boldsymbol\sigma)^2\rangle\sim \langle \boldsymbol\sigma^2\rangle\langle(\partial_\mu\boldsymbol\sigma)^2\rangle$ is not the case.
After some manipulations for the one-loop contribution to this condensate we have:
$$
\langle (\boldsymbol\sigma\cdot\partial_\mu\boldsymbol\sigma)^2\rangle=-i\hbar\left[ G^{ab}_{ii}\partial_\mu\sigma^a_i\partial_\mu\sigma^b_{i}+\left.\sigma^a_i\sigma^b_{i}\partial_\mu^k\partial_\mu^l G^{ab}_{kl}\right|_{k=l=i}+\left.\partial_\mu^i\sigma^a_i\sigma^b_i\partial_\mu^k G^{ab}_{ik}\right|_{k=i}+\left.\partial_\mu^i\sigma^b_i\sigma^a_i\partial_\mu^l G^{ab}_{li}\right|_{l=i}\right].
$$
It is easy to verify that:
$$
G^{ab}_{ii}\partial_\mu\sigma^a_i\partial_\mu\sigma^b_{i}=\frac{Nm^2}{\gamma}D^{-1}_{0ii};\quad \left.\partial_\mu^i\sigma^b_i\sigma^a_i\partial_\mu^l G^{ab}_{li}\right|_{l=i}=\left.\partial_\mu^i\sigma^a_i\sigma^b_i\partial_\mu^k G^{ab}_{ik}\right|_{k=i}= \left.\sigma^a_i\sigma^b_{i}\partial_\mu^k\partial_\mu^l G^{ab}_{kl}\right|_{k=l=i}=0.
$$
Consequently
$$
\langle (\boldsymbol\sigma\cdot\partial_\mu\boldsymbol\sigma)^2\rangle=-i\hbar\frac{Nm^2}{\gamma}D^{-1}_{0ii}=\frac{\hbar}{4\pi}\frac{Nm^2}{\gamma}N_\varepsilon\neq \langle \boldsymbol\sigma^2\rangle\langle(\partial_\mu\boldsymbol\sigma)^2\rangle.
$$
The last condensate  we would like to calculate (in one-loop approximation) is $\langle\boldsymbol\sigma^2(\partial_\mu\boldsymbol\sigma)^2\rangle.$
The result will be presented
without detailed calculations:
$$
\langle\boldsymbol\sigma^2(\partial_\mu\boldsymbol\sigma)^2\rangle=\boldsymbol\sigma^2_0\left((\partial_\mu\boldsymbol\sigma)^2_0+2\hbar\partial_\mu\boldsymbol\sigma_0\partial_\mu\boldsymbol\sigma_1\right)-i\hbar(\partial_\mu\boldsymbol\sigma_0)^2(N-1)D^{-1}_{0ii}-\left.i\hbar\boldsymbol\sigma^2_0\partial_\mu^k\partial_\mu^lG^{aa}_{kl}\right|_{k=l=i}=
$$
$$
=\langle\boldsymbol\sigma^2\rangle\langle(\partial_\mu\boldsymbol\sigma)^2\rangle+\hbar\frac{Nm^2}{4\pi\gamma}(N-1)N_\varepsilon.
$$
This condensate is not factorizes too.

\section{Conclusion and Discussion}
Effective action is central quantity in quantized fields theory because it contains all the quantum information about any quantum field model.
As it follows from the presented here consideration the system of equations for $\Gamma$  and especially for  $W$ allow to calculate directly various  vacuum expectations.

It should be  underlined once more that our aim is to demonstrate a method of calculation of the vacuum expectations, therefore some of the obtained    results  above are not new ones.
Here it is interesting to compare our results with those obtained by the $1/N$ expansion method in \cite{nsvz}, \cite{ksen}. It is not surprising when results obtained
by different approximations are different.
In \cite{ksen} and  \cite{nsvz}
the relation $\boldsymbol{\sigma}_0\cdot\boldsymbol{\sigma}_1=0$  (in our notations)
which was derived in the frame of the $1/N$ expansion is used.
But in our calculations we have $\boldsymbol{\sigma}_0\cdot\boldsymbol{\sigma}_1=i(N-1)D^{-1}_{0ii}/2$, this is  based on  relation Eq.(\ref{sig0sig1}) which is derived as first order contribution to quasiclassical expansion of the exact quantum equations (\ref{phisigmaW}).
As a result, the vacuum expectation of the trace of the energy-momentum tensor
in \cite{ksen} is expressed through $\boldsymbol{\sigma}_1\cdot\boldsymbol{\sigma}_1$ and $\partial_\mu\boldsymbol{\sigma}_1\cdot\partial^\mu\boldsymbol{\sigma}_1$ terms. This corresponds not to the first order quantum corrections, but
 to the second order ones. In $1/N$ approximation factorization properties of many condensates are widely used \cite{nsvz}, from results of our last section follows
 that in quasi classical expansion  not in all cases factorization takes place.

 In this article the role of classical solutions in quantum corrections are presented explicitly.

It seems  equations for effective action will be more effective in studying of nonperturbative structure of quantum fields, especially the weak and strong coupling limits.
As was mentioned above the weak coupling limit in any quantum field model is singular one. It will be very interesting to investigate this singularity.
Some examples of such investigation will be presented soon.

\bb
\bibitem{nsvz}
V.A.Novikov, M.A.Shifman, A.I.Vainshtein and V.I.Zakharov, Two dimensional sigma models: modelling  nonperturbative effects in QCD, Phys.Rep.\textbf{116}(1984)103-171.
\bibitem{pol} A.M.Polyakov, Interaction of goldstone particles in two dimensions, Phys.Lett.{\bf B 59}(1975)79-81.
\bibitem{ksen} V.G.Ksenzov, On the calculation of the vacuum energy density in sigma models, Phys.Lett.{\bf B 367}(1996)237-241.
\bibitem{fm}
B.A.Fayzullaev and M.M.Musakhanov, Two-loop effective action for theories with fermions, Annals of Phys.(NY) {\bf 241}
(1995)394.
\bibitem{dw} B.S.DeWitt, Dynamical Theory of Groups and Fields,
Gordon and Breach, New York, 1965.
\bibitem{fbg}B.A.Fayzullaev, S.Garnov and D.V.Galkin, in the Proc. of X-Int.
Conference "Problems of Quantum Field Theory", Alushta, Ed.by
D.Shirkov,
D.Kazakov and A.Vladimirov (1996)174-177;\\
B.A.Fayzullaev, S.V.Bondarenko and D.V.Galkin, "Total two-loop
effective action for QED", in the Proc.of VII-Int.Conference on
Mathematical Physics - Caspian Conference - 1995, Ed.by F.Ardalan,
H.Arfaei and S.Ruhani, Tehran, IPM (1997)15-24.
\bibitem{polbel}  A.A.Belavin  and A.M.Polyakov, Metastable states of two-dimensional isotropic ferromagnetic,  JETP Letters v.22(1975)245.
\bibitem{sf} A.A.Slavnov and L.D.Faddeev, Gauge fields. An introduction to quantum theory, Nauka, Moscow, 1988 (in russian).

\eb
\end{document}